\documentstyle[aps,prb,preprint]{revtex}
\begin{document}
\draft
\tightenlines
\title{Dispersion of the strongly correlated electron system \\
near the Fermi level}
\author{A.~Sherman}
\address{Institute of Physics, University of Tartu, Riia
142, 51014 Tartu, Estonia}
\date{\today}
\maketitle
\begin{abstract}
The reason for the appearance of two distinct bands in the photoemission
of the single CuO$_2$ plane Bi$_2$Sr$_{2-x}$La$_x$CuO$_{6+\delta}$ near
the Fermi level is discussed on the basis of the self-consistent
solution of the two-dimensional $t$-$J$ model.
\end{abstract}


Angle resolved photoemission spectroscopy data have been one of the most
important sources of information about the electronic structure of the
high temperature superconductors (see, e.g., Ref.~\onlinecite{Shen} and
references therein). One of the most interesting results obtained
recently with this experimental method is the observation of two
distinct emissions near the Fermi energy on
Bi$_2$Sr$_{2-x}$La$_x$CuO$_{6+\delta}$ in the vicinity of the $(\pi,0)$
point. \cite{Manzke,Janowitz} This crystal is a single CuO$_2$ layer
material with sufficiently decoupled layers. Therefore the two emissions
cannot be ascribed to the bilayer splitting. 

Normal state photoemission spectra obtained in
Ref.~\onlinecite{Janowitz} in the optimally doped crystal with
$T_c=29$~K are reproduced in Fig.~\ref{figi}(a)--(c). The measurements
were performed at the temperature $T=45$~K. Along the $\Gamma M$
direction [${\bf k}_M=(\pi, 0)$] the spectra consist of two emission
bands which are well resolved for wave vectors some distance away from
the $M$ point and merge to a structure which cannot be resolved into
individual contributions on approaching this point. The emission band
with the higher binding energy has a considerably larger halfwidth and
stronger dispersion in comparison with the band with the lower binding
energy.

In this Communication I would like to call attention to the fact that
in many respects similar spectra were obtained
\cite{sherman98,sherman94} in the two-dimensional $t$-$J$ model of the
CuO$_2$ plane. In these works self-consistent calculations of the hole
and magnon Green's functions were carried out for the cases from light
to moderate doping with the use of the Born and different versions of
the spin-wave approximations. As pointed out in these works, for moderate
hole concentrations $x$ the hole spectrum contains two dispersive
features for wave vectors near the boundary of the magnetic Brillouin
zone. One of these features -- a narrow intensive peak slightly below
(in the electron picture) the Fermi level -- is connected with the
so-called spin-polaron band. In a lightly doped crystal the width of
this band is of the order of the exchange constant $J$ which is much
smaller than the hopping constant $t$ of the model for parameters of
cuprate perovskites. The spin-polaron bandwidth is characterized by the
parameter of magnetic excitations because on the antiferromagnetic
background the hole movement is accompanied by the magnon absorption and
emission. The short-range antiferromagnetic ordering is retained in
moderately doped crystals. In these conditions a part of the
spin-polaron band is preserved near the boundary of the magnetic
Brillouin zone. The second dispersive feature is a broad maximum which
dispersion is characterized by the second energy parameter of the model
$t$. The shape of this dispersion reproduces with some distortion the
shape of the two-dimensional nearest-neighbor band.
\cite{sherman98,sherman94} 

An example of the hole spectral function $A$ -- the imaginary part of
the hole Green's function -- with the two mentioned dispersive features
is shown in Fig.~\ref{figii} for the parameters $t=0.5$~eV, $J=0.1$~eV,
$x\approx 0.1$ and $T=0$. The calculations \cite{sherman98} were carried
out on a 20$\times$20 lattice with the use of the spin-wave
approximation \cite{takahashi} modified for short-range order. The
narrow spin-polaron peak which intensity grows on approaching the $M$
point is located near the Fermi level (its energy is taken as zero). For
wave vectors near the $\Gamma$ point two broad maxima, one below and one
above the Fermi level, are seen in Fig.~\ref{figii}. Both of them belong
to the dispersion with the characteristic energy $t$ and appear together
in the spectrum due to the broken symmetry implied in the spin-wave
approximation. In the used magnetic Brillouin zone, which is half as
much as the usual one, the spectral function contains features
corresponding to two points of the usual Brillouin zone. These points
are connected by the antiferromagnetic wave vector $(\pi,\pi)$.

The experimental photoemission spectrum can be modelled by the above
spectral function convoluted with the Gaussian of an appropriate width
for the experimental resolution and cut off by the Fermi distribution
function. The spectra obtained in this way for three wave vectors are
shown in the right panel of Fig.~\ref{figi}. In accord with the
experimental conditions of Ref.~\onlinecite{Janowitz} in these
calculations the width of the Gaussian and the temperature in the Fermi
distribution were set equal to 20~meV and 45~K, respectively. In spite
of some differences in relative intensities and widths of the spectral
features, the calculated photoemission spectrum has the same structure
as the experimental spectrum \cite{Janowitz} with two spectral features
(only the foot of the spin-polaron peak is shown on the right panel of
Fig.~\ref{figi} to demonstrate more clearly the maximum or shoulder with
a larger binding energy and smaller peak intensity). The experimental
and calculated dispersions of the two spectral features are compared in
Fig.~\ref{figiii}. As is seen from the figure, the experimental and
calculated dispersions are similar and the binding energies are of the
same order of magnitude.

In the experiment, the splitting of the photoemission band was observed
along the $\Gamma M$ direction and was not detected near the
$(\pi/2,\pi/2)$ point. \cite{Janowitz} The comparison of the calculated
photoemission spectra for ${\bf k}=(0.6\pi,0)$ and $(\pi/2,\pi/2)$ is
given in Fig.~\ref{figiv}. The spectra were obtained in the same way and
for the same parameters as for the right panel of Fig.~\ref{figi}. As
can be seen from Fig.~\ref{figiv}, the broad spectral feature reveals
itself much more clearly at $(0.6\pi,0)$. This is connected with the
fact that near $(\pi/2,\pi/2)$ central frequencies of the spin-polaron
peak and the broad maximum are close. In this region of the Brillouin
zone the spin-polaron peak touches the Fermi level and the broad maximum
crosses it. The Fermi surface of the 2D $t$-$J$ model is shown in
Fig.~\ref{figv} for the underdoped case. \cite{sherman94}

Recently \cite{Chuang} two dispersive features were resolved near the
Fermi level in photoemission of the underdoped and optimally doped
Bi$_2$Sr$_2$CaCu$_2$O$_{8+\delta}$ with two CuO$_2$ planes in the unit
cell. \cite{Feng} In Ref.~\onlinecite{Chuang} this splitting of the
photoemission band was connected with a coupling between adjacent
CuO$_2$ planes. Since energy distribution curves and values of the band
splitting in that work are similar to those observed in the single
CuO$_2$ plane Bi$_2$Sr$_{2-x}$La$_x$CuO$_{6+\delta}$, another possible
mechanism of the band splitting in Bi$_2$Sr$_2$CaCu$_2$O$_{8+\delta}$ --
the above-discussed peculiarity of the spectrum of the 2D strongly
correlated system -- has to be also taken into account. It counts in
favour of this latter mechanism that the Fermi surface obtained in
Ref.~\onlinecite{Chuang} (see Fig.~1g there) is very close to the Fermi
surface in Fig.~\ref{figv}.

In summary, it was demonstrated that the two dispersive features in the
hole spectral function of the two-dimensional $t$-$J$ model are similar
to the features observed in the photoemission spectrum of the single
CuO$_2$ plane Bi$_2$Sr$_{2-x}$La$_x$CuO$_{6+\delta}$. Thus, this
peculiarity of the spectrum of the strongly correlated system can be
considered as a possible reason for the photoemission band splitting in
this crystal.

\begin{figure}\caption{Left panel: Normal state photoemission spectra
of\protect\nolinebreak\ the optimally doped
Bi$_2$Sr$_{2-x}$La$_x$CuO$_{6+\delta}$ ($T_c=29$~K, $T=45$~K) for ${\bf
k}=(0.83\pi,0)$ (a), $(0.69\pi,0)$ (b), and $(0.62\pi,0)$ (c).
\protect\cite{Janowitz} Right panel: Normal state photoemission spectra
in the 2D $t$-$J$ model for ${\bf k}=(0.8\pi,0)$ (d), $(0.7\pi,0)$ (e),
and $(0.6\pi,0)$ (f). The spectra were obtained from spectral functions
self-consistently calculated \protect\cite{sherman98} for the hole
concentration $x \approx 0.1$ in a 20$\times$20 lattice. The spectral
functions were convoluted with Gaussian with the width of 20~meV to
model experimental resolution and cut off with the Fermi distribution
with $T=45$~K. The hopping and exchange parameters of the model are
$t=0.5$~eV and $J=0.1$~eV.
}
\label{figi}\end{figure}

\begin{figure}\caption{The hole spectral function of the 2D $t$-$J$
model along the $(0,0)-(\pi,0)$ direction of the Brillouin zone. The
self-consistent calculations \protect\cite{sherman98} were carried out
on a 20$\times$20 lattice for the parameters $t=0.5$~eV, $J=0.1$~eV,
$x\approx 0.1$ and $T=0$. The Fermi energy is taken as zero of energy.
}
\label{figii}\end{figure}

\begin{figure}\caption{(a) The dispersions of the two features in the
photoemission spectrum of Bi$_2$Sr$_{2-x}$La$_x$CuO$_{6+\delta}$ (the
dispersions were derived in Ref.~\protect\onlinecite{Janowitz} by
fitting the experimental spectra by two Lorentzians convoluted with a
Gaussian to imitate the experimental resolution and cut off by the Fermi
distribution). (b) The dispersions of the maxima in the spectral
function in Fig.~\protect\ref{figii}.
} 
\label{figiii}\end{figure}

\begin{figure}\caption{The photoemission spectra in the $t$-$J$ model
for ${\bf k}=(0.6\pi,0)$ (solid line) and $(\pi/2,\pi/2)$ (dashed line).
The spectra were obtained in the same way and for the same parameters as
for the right panel of Fig.~\protect\ref{figi}.
} 
\label{figiv}\end{figure}

\begin{figure}\caption{The Fermi surface of the 2D $t$-$J$ model for the
underdoped case. \protect\cite{sherman94} The solid segments along the
boundary of the magnetic Brillouin zone are formed by points where the
spin-polaron band touches the Fermi level. It is crossed by the broader
dispersive feature along the dashed curves.
} 
\label{figv}\end{figure}

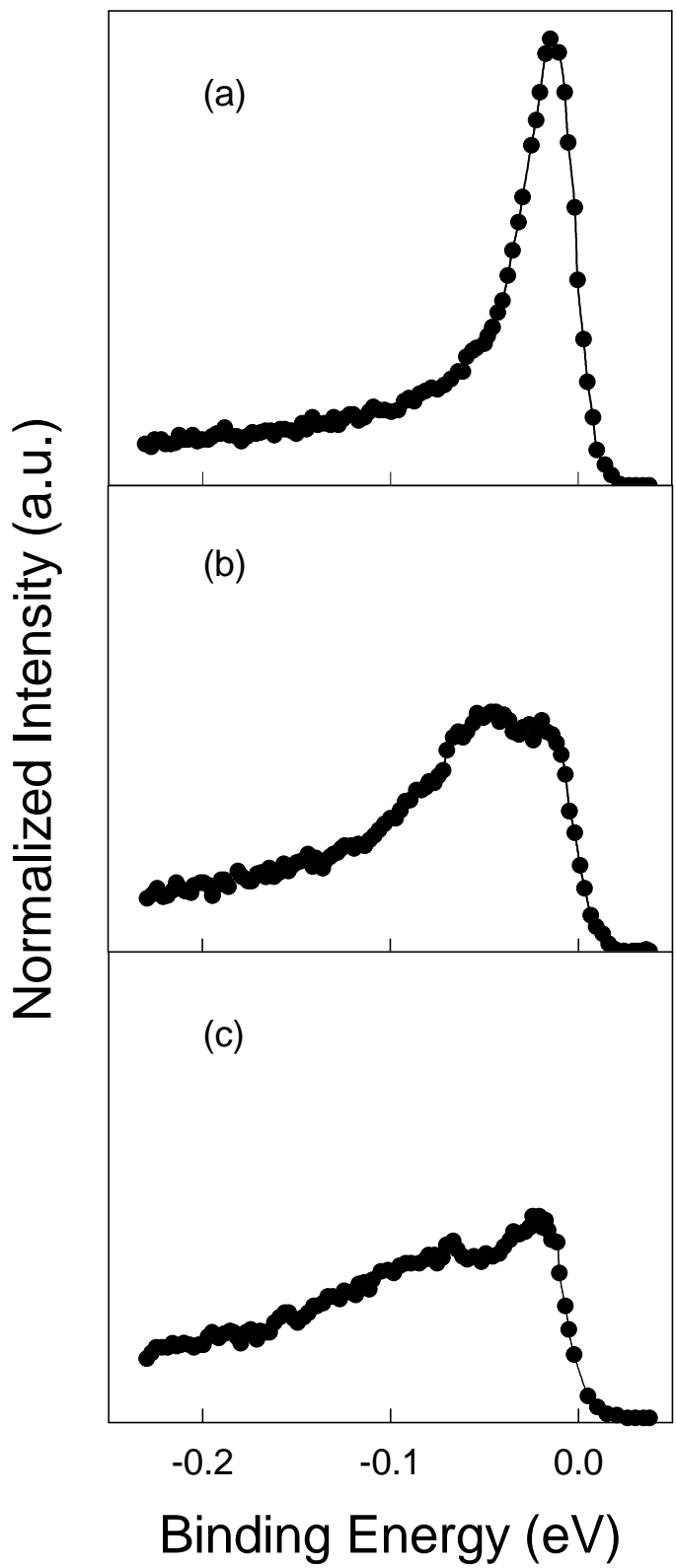
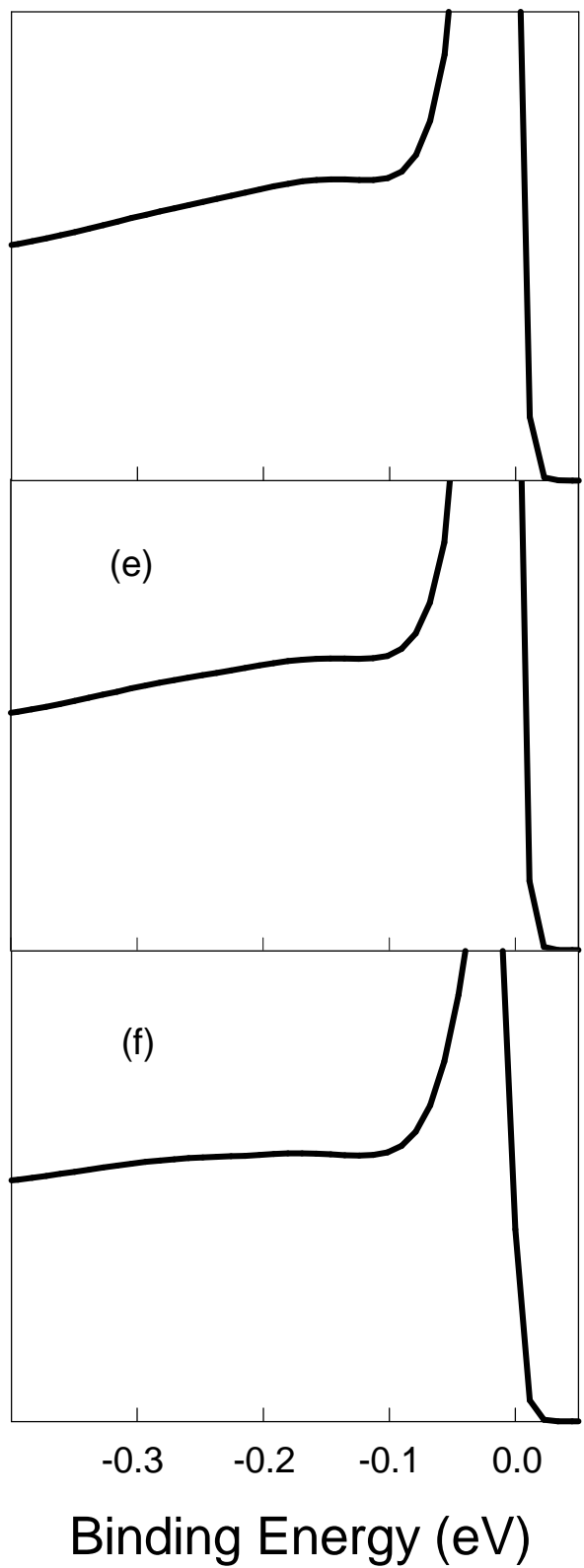
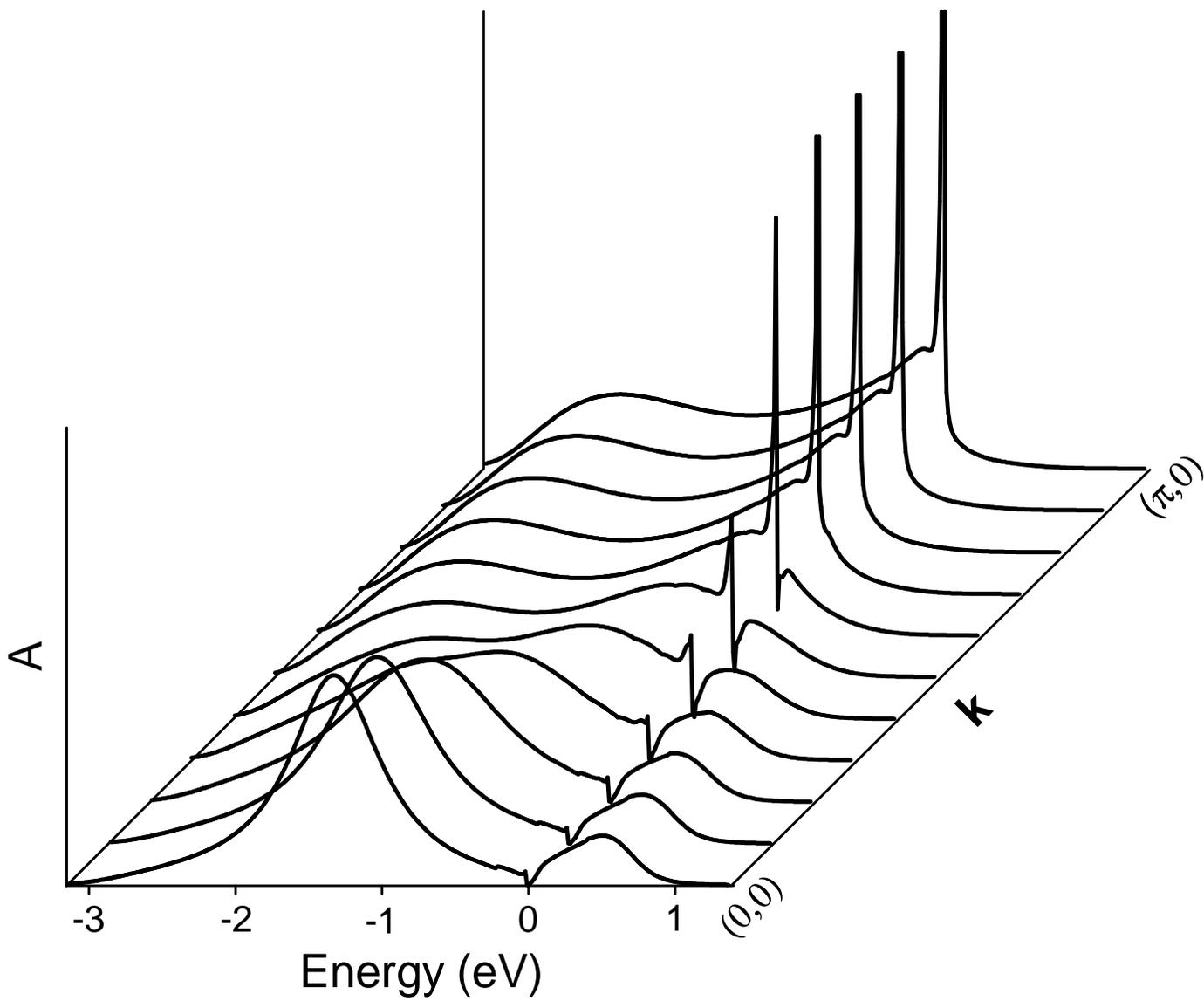
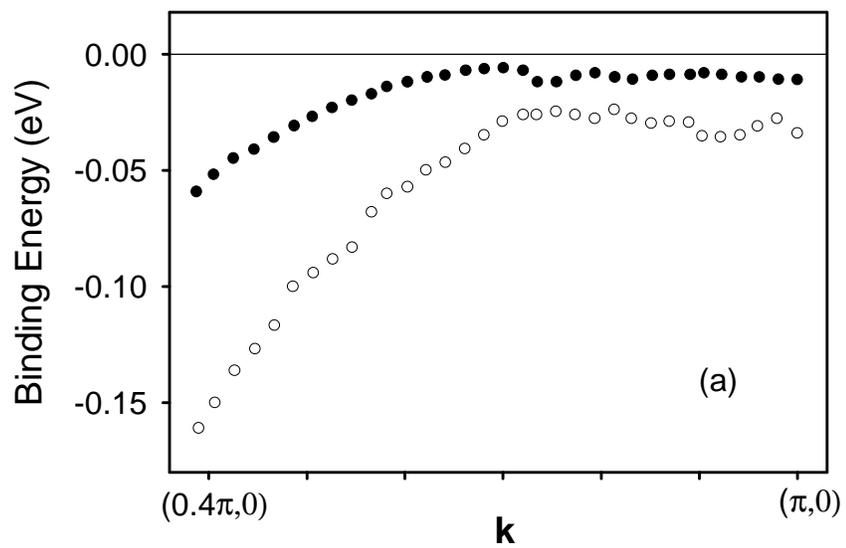
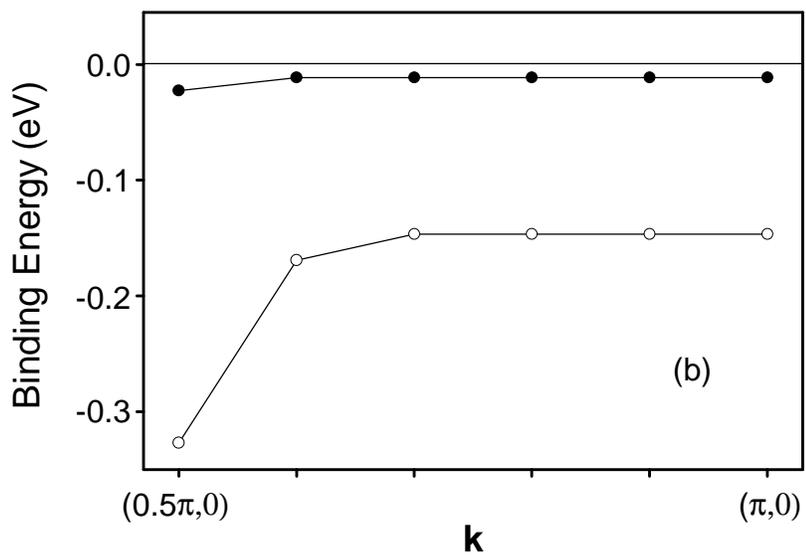
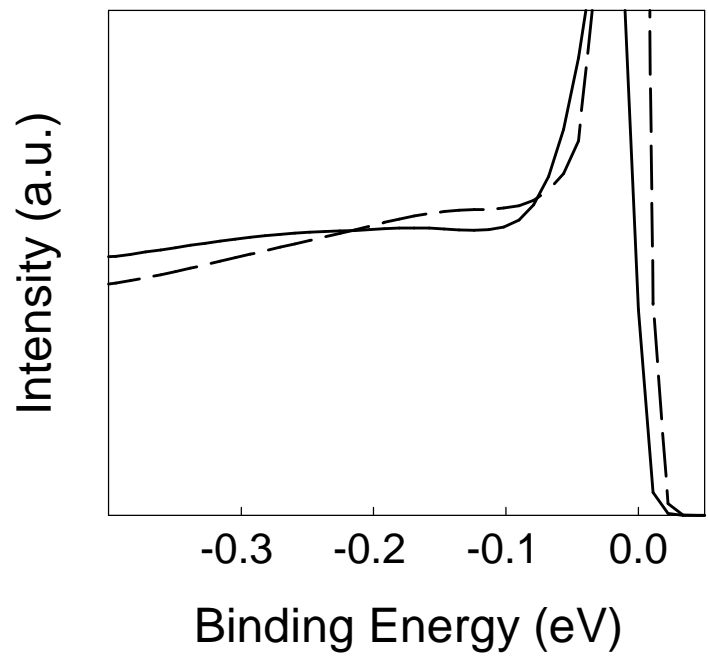
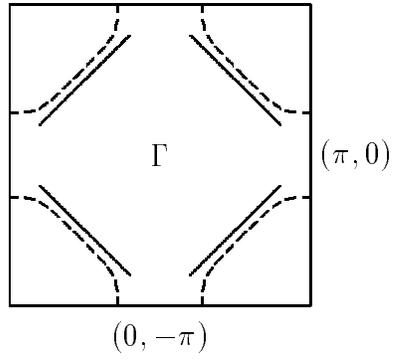
\end{document}